\documentclass[5p]{elsarticle}
\usepackage{amssymb}
\usepackage{amsmath}
\usepackage{color}
\usepackage{graphicx}
\usepackage[english]{babel}

\newcommand{\er}{$\pm$}
\newcommand{\BR}{\ensuremath{\mathrm{BR}}}

\begin{document}

\begin{frontmatter}

\title{{\bf\boldmath  New data on $\vec{\gamma} \vec{p}\rightarrow \eta p$ with polarized photons and protons 
and their implications for $N^* \to N\eta$ decays}}


\author[label1]{J.~M\"uller}
\author[label1]{J.~Hartmann}
\author[label1]{M.~Gr\"uner}
\author[label1]{F.~Afzal}
\author[label1,label2]{A.V.~Anisovich}
\author[label3]{B.~Bantes}
\author[label1,label2]{D.~Bayadilov}
\author[label1]{R.~Beck}
\author[label1]{M.~Becker}
\author[label2]{Y.~Beloglazov}
\author[label4]{M.~Berlin}
\author[label4]{M.~Bichow}
\author[label1]{S.~B\"ose}
\author[label1,label5]{K.-T.~Brinkmann}
\author[label6]{T.~Challand}
\author[label7]{V.~Crede}
\author[label5]{F.~Dietz}
\author[label6]{M.~Dieterle}
\author[label5]{P.~Drexler}
\author[label3]{H.~Dutz}
\author[label3]{H.~Eberhardt}
\author[label3]{D.~Elsner}
\author[label3]{R.~Ewald}
\author[label3]{K.~Fornet-Ponse}
\author[label5]{S.~Friedrich}
\author[label3]{F.~Frommberger}
\author[label1]{C.~Funke}
\author[label1]{M.~Gottschall}
\author[label2]{A.~Gridnev}
\author[label3]{S.~Goertz}
\author[label1,label5]{E.~Gutz}
\author[label1]{C.~Hammann}
\author[label5]{V.~Hannen}
\author[label3]{J.~Hannappel}
\author[label4]{J.~Herick}
\author[label3]{W.~Hillert}
\author[label1]{P.~Hoffmeister}
\author[label1]{C.~Honisch}
\author[label3]{O.~Jahn}
\author[label3]{T.~Jude}
\author[label6]{I.~Jaegle}
\author[label6]{A.~K\"aser}
\author[label1]{D.~Kaiser}
\author[label1]{H.~Kalinowsky}
\author[label1]{F.~Kalischewski}
\author[label3]{S.~Kammer}
\author[label6]{I.~Keshelashvili}
\author[label1]{P.~Klassen}
\author[label3]{V.~Kleber}
\author[label3]{F.~Klein}
\author[label1]{E.~Klempt}
\author[label1]{K.~Koop}
\author[label6]{B.~Krusche}
\author[label1]{M.~Kube}
\author[label1]{M.~Lang}
\author[label2]{I.~Lopatin}
\author[label6]{Y.~Maghrbi}
\author[label1]{P.~Mahlberg}
\author[label5]{K.~Makonyi}
\author[label3]{F.~Messi}
\author[label5]{V.~Metag}
\author[label4]{W.~Meyer}
\author[label1]{J.~M\"ullers}
\author[label5]{M.~Nanova}
\author[label1,label2]{V.~Nikonov}
\author[label2]{D.~Novinski}
\author[label5]{R.~Novotny}
\author[label1]{D.~Piontek}
\author[label4]{G.~Reicherz}
\author[label1]{C.~Rosenbaum}
\author[label6]{T.~Rostomyan}
\author[label4]{B.~Roth}
\author[label1,label2]{A.~Sarantsev}
\author[label1]{C.~Schmidt}
\author[label3]{H.~Schmieden}
\author[label1]{R.~Schmitz}
\author[label1]{T.~Seifen}
\author[label1]{V.~Sokhoyan}
\author[label1]{A.~Thiel}
\author[label1]{U.~Thoma\corref{cor1}}
\ead{thoma@hiskp.uni-bonn.de}
\author[label1]{M.~Urban}
\author[label1]{H.~van~Pee}
\author[label1]{D.~Walther}
\author[label1]{C.~Wendel}
\author[label4]{U.~Wiedner}
\author[label1,label7]{A.~Wilson}
\author[label1]{A.~Winnebeck}
\author[label6]{L.~Witthauer}
\author{\\[2ex]CBELSA/TAPS Collaboration}

\cortext[cor1]{Corresponding author.}

\address[label1]{Helmholtz--Institut f\"ur Strahlen-- und Kernphysik, Universit\"at Bonn, Germany}
\address[label2]{National Research Centre "Kurchatov Institute", 
Petersburg Nuclear Physics Institute, Gatchina, Russia}
\address[label3]{Physikalisches Institut, Universit\"at Bonn, Germany}
\address[label4]{Institut f\"ur Experimentalphysik I, Ruhr--Universit\"at Bochum, Germany}
\address[label5]{II.~Physikalisches Institut, Universit\"at Gie{\ss}en, Germany}
\address[label6]{Physikalisches Institut, Universit\"at Basel, Switzerland}
\address[label7]{Department of Physics, Florida State University, Tallahassee, FL 32306, USA}

\begin{abstract}
The polarization observables $T, E, P, H$, and $G$ in photoproduction of $\eta$ mesons off 
protons are measured for photon energies from threshold to $W=2400\,$MeV ($T$), 2280\,MeV ($E$),
1620\,MeV ($P, H$), or 1820\,MeV ($G$), covering nearly the full solid angle. 
The data are compared to predictions from the SAID, MAID, J\"uBo, 
and BnGa partial-wave analyses. A refit within the BnGa approach including further data yields 
precise branching ratios for the $N\eta$ decay of nucleon resonances. 
A $N\eta$-branching ratio of $0.33\pm 0.04$ for $N(1650)1/2^-$ is found, which reduces the large and controversially 
discussed $N\eta$-branching ratio difference of the two lowest mass $J^P=1/2^-$-resonances 
significantly. 
\end{abstract}

\begin{keyword}
baryon spectroscopy  \sep meson photoproduction \sep polarization observables 
\end{keyword}

\end{frontmatter}

\section{Introduction}
The properties of excited states of protons and neutrons, their masses, widths and decays, reflect their internal dynamics. 
Quark models describe the excitation spectrum of nucleons by the interaction of three constituent 
quarks in a confinement potential adding a residual interaction such as one-gluon~\cite{Isgur:1977ef,Capstick:1986bm} 
or pseudoscalar-meson~\cite{Glozman:1995fu} exchange, or instanton induced interactions~\cite{Loring:2001kx}. 
QCD calculations on the lattice~\cite{Edwards:2011jj} -- even though using unphysically large quark masses -- yield a similar pattern. 
A very different view assumes that quarks and gluons are 
not the appropriate degrees of freedom to describe nucleon resonances; instead, resonances are generated dynamically 
from their hadronic decay products~\cite{Nacher:1999vg,Kolomeitsev:2003kt,Bruns:2010sv,Mai:2012wy}. 
Properties of baryon resonances differentiating between the models are of particular importance.

The surprising decay pattern of the two lowest-mass nucleon excitations, $N(1535)1/2^-$ and $N(1650)1/2^-$ with 
spin-parity $J^P{=}1/2^-$ and carrying an orbital angular momentum $L{=}1$, has always been a challenge for model builders.
In 2010, the $N\eta$ branching ratio of $N(1535)1/2^-$ was estimated by the Particle Data Group \cite{PDG2010} 
to 45--60\%, and only 3--10\% for $N(1650)1/2^-$. 
Several interpretations have been offered to explain the unexpectedly large\linebreak $N(1535)1/2^- \to N\eta$ branching ratio:

i) Within the quark model \cite{Isgur:1977ef}, the one-gluon exchange interaction leads to a mixing angle 
of the two states with defined total quark spin $S$, $|J{=}1/2;L{=}1,S{=}1/2\rangle$ and \linebreak$|J{=}1/2;L{=}1,S{=}3/2\rangle$. 
At this mixing angle, the higher-mass state $N(1650)1/2^-$ nearly decouples from $N\eta$; the lower-mass state 
$N(1535)1/2^-$ acquires a large $N\eta$ branching ratio.\\ 
ii) In the quark model \cite{Glozman:1995tb}, the large $N\eta$ branching ratio is explained as a consequence of 
a dynamical clusterization into a quark-diquark configuration.\\
iii) The low mass of the $N(1440)1/2^+$ Roper resonance and the large $N(1535)1/2^-\to N\eta$ coupling may both be explained by 
large pentaquark components in their wave functions~\cite{Zou:2007mk}.\\
iv) In \cite{Mai:2012wy,Kaiser:1995cy}, 
$N(1535)1/2^-$ is generated dynamically and interpreted as quasi-bound $K\varSigma$-$K\varLambda$-state decaying strong\-ly 
into $N\eta$ via coupled-channel effects. 

All models agree on the conclusion -- driven by experimental information -- that the $N(1535)1/2^-\to N\eta$
branching ratio is much larger than that for $N(1650)1/2^-$$\to$ $N\eta$ decays. These results were, 
however, derived from rather poor data on $\pi^-p\to \eta n$ and from differential cross sections and the beam 
asymmetry for $\gamma p\to \eta p$. Neither data set fully constrains the amplitudes 
governing pion- or photo-production of $\eta$ mesons. Thus, a wide range of results on the $N\eta$ 
branching ratio was reported in the literature.
 
Vrana {\it et al.}~\cite{Vrana:1999nt} fitted data on $\pi N$ inelastic reactions, with
$\pi N$, $\eta N$ and $\pi\pi N$ as admitted final states. When these three final states were included, 
$N\eta$ branching ratios for $N(1650)1/2^-$ of 16\%, 25\%, and 6\% were derived using different model 
assumptions. The last model was considered to be the best one, and a branching ratio $\BR=0.06\pm0.01$ was quoted as final 
result. The authors pointed out that the data on $\pi^-p\to\eta n$ is both, limited and of uncertain 
quality. This statement holds true, of course, for all analyses using those data.   

Penner and Mosel performed a coupled-channel analysis of a large number of reactions. The authors gave a 
branching ratio between $0.004$ and $0.051$ \cite{Penner:2002ma,Penner:2002md}.
Shklyar {\it et al.} \cite{Shklyar:2012js} gave a branching of $0.01\pm0.02$, which we read as $<0.03$. 

The Bonn-Gatchina group \cite{Anisovich:2011fc} reported a value of $0.18\pm0.04$ from a study of a
large body of pion and photo-induced reactions, a value that superseded an earlier fit \cite{Thoma:2007bm} 
to a smaller data sample reporting $0.15\pm0.06$. 

A new $\eta$MAID2017-solution~\cite{Kashevarov:2017kqb} including the data~\cite{Akondi:2014ttg,Senderovich:2015lek,Crede:2003ax,Bartalini:2007fg,Williams:2009yj} finds $0.28\pm0.11$ using an $A_{1/2}$-value of $+0.045$. MAID2018 reports a branching ratio of $0.19\pm0.06$, using $A_{1/2}=+0.055$~\cite{MAID2018}. Both solutions lead to the same $A_{1/2}\sqrt{\BR(N\eta)}$ for $N(1650)1/2^-$.  

Shrestha and Manley \cite{Shrestha:2012ep} performed coupled-channel fits to pion-induced reactions and 
determined a $N\eta$ branching ratio for $N(1650)1/2^-$ of $0.21\pm0.02$ where the error is of 
statistical nature. Batinic {\it et al.} fitted data on the reactions $\pi N\to \pi N$ and $\eta N$ and obtained
$0.13\pm0.05$.

Tryasuchev \cite{A.Tryasuchev:2014jda} introduces a quantity $\xi$ defined as 
\begin{eqnarray}
\label{eq:trya}
\xi_{1/2,\,3/2} &=& \sqrt{\frac{k\, M_p\, \BR_\eta}{q\, M_{R}\, \varGamma_{R}}}\cdot A_{1/2,\,3/2}
\end{eqnarray}
with the proton mass $M_p$ and the resonance mass $M_{R}$ and width $\varGamma_{R}$. 
$k$ and $q$ are the decay momenta of photon or $\eta$ in the center-of-mass system.
In a fit to data on $\eta$ photoproduction, Tryasuchev finds
for $N(1650)1/2^-$ the value $\xi_{1/2}=0.0975$\,GeV$^{-1}$, from which we deduce
 $A_{1/2} \cdot \sqrt{\BR_\eta} =0.034$\,GeV$^{-1/2}$ (using PDG-values~\cite{PDG2019} for $M_{R} \varGamma_{R}$). 
%
%
No error is given in 
\cite{A.Tryasuchev:2014jda} for $\xi$. With this value for $\xi$ and our value for  $A_{1/2}$ reported below, 
the $N\eta$ branching ratio should be in the order of 1. 

The analysis of 
$\eta$ production in pion and photo-induced reactions by the 
J\"ulich/Bonn group~\cite{deborah} finds a $N\eta$ branching ratio 
that is more than a factor six larger for $N(1535)1/2^-$ than for $N(1650)1/2^-$.

In this letter, we present results from a study of 
$\gamma p\to \eta p$ using a longitudinally 
or transversely polarized target with polarization $p_T$ and linearly or circularly polarized photons with polarization 
$p_\gamma$ or $p_{\odot}$, respectively. 
The results of these different data sets are presented here in a single letter since we believe that only the complete 
information can constrain a partial-wave analysis sufficiently well to lead to unambiguous results on $N^*\to N\eta$ decays. 
For details on the measurements and data analyses, we quote earlier publications on $\gamma p\to \pi^0 p$ on $E$~\cite{Gottschall:2013uha,Gottschall:2019pwo}, $T,P,H$ \cite{Hartmann:2014mya,Hartmann:2015kpa}, and $G$~\cite{Thiel:2012yj,Thiel:2016chx}.

\section{The experiment}

\subsection{Experimental setup}
The experiment was carried out at the ELectron Stretch\-er Accelerator ELSA in Bonn~\cite{Hillert:2006yb}. 
Photons with circular polarization $p_{\odot}$ were produced by scattering a 2.335\,GeV beam of longitudinally polarized 
electrons off a brems\-strahlung target: $p_{\odot}$ decreases from 0.63 at the maximal tagged photon energy of 
2.29~GeV to  0.34 at 1\,GeV. 
Linearly polarized photons with polarization  $p_\gamma$ stem from coherent brems\-trahlung of 3.2\,GeV electrons 
off an aligned diamond. For the  measurement of $T, P$, and $H$, the coherent edge of the crystal was set to achieve 
a maximum polarization of  $p_\gamma$ = 65\% at 850\,MeV. 
For $G$, three polarization settings were used. Here, the maximum linear polarization reached was 65\% at 
860\,MeV, 59\% at 1050\,MeV, and 55\% at 1270\,MeV.

The electrons passed through a magnet hitting a tagging hodoscope, which defined the energy of the brems\-strahlung 
photons. The photon beam impinged on the Bonn frozen spin butanol ($\rm C_4H_9OH$) target containing either longitudinally or transversely 
polarized protons~\cite{Dutz:2004eb}. 
The target was surrounded by a three-layer scintillation fiber detector~\cite{Suft:2005cq} used 
for the identification of charged particles and by the Crystal Barrel electromagnetic calorimeter~\cite{CB} consisting of 
1230 CsI(Tl)-crystals. In the forward direction below polar angles of 30$^\circ$, two further calorimeters, 
the forward detector consisting of 90 CsI(Tl)-crystals and the forward TAPS-wall~\cite{TAPS} (216 BaF$_2$ crystals), 
provided calorimetric information. Plastic scintillators in front of the forward crystals  
allowed for the identification of charged particles. 
A $\rm CO_2$ Cherenkov detector placed before the forward TAPS-wall vetoed signals due to electron or positron hits, which are due 
to electromagnetic background produced in the target. 
\begin{figure*}[t!]\begin{center}
\includegraphics[width=0.87\textwidth]{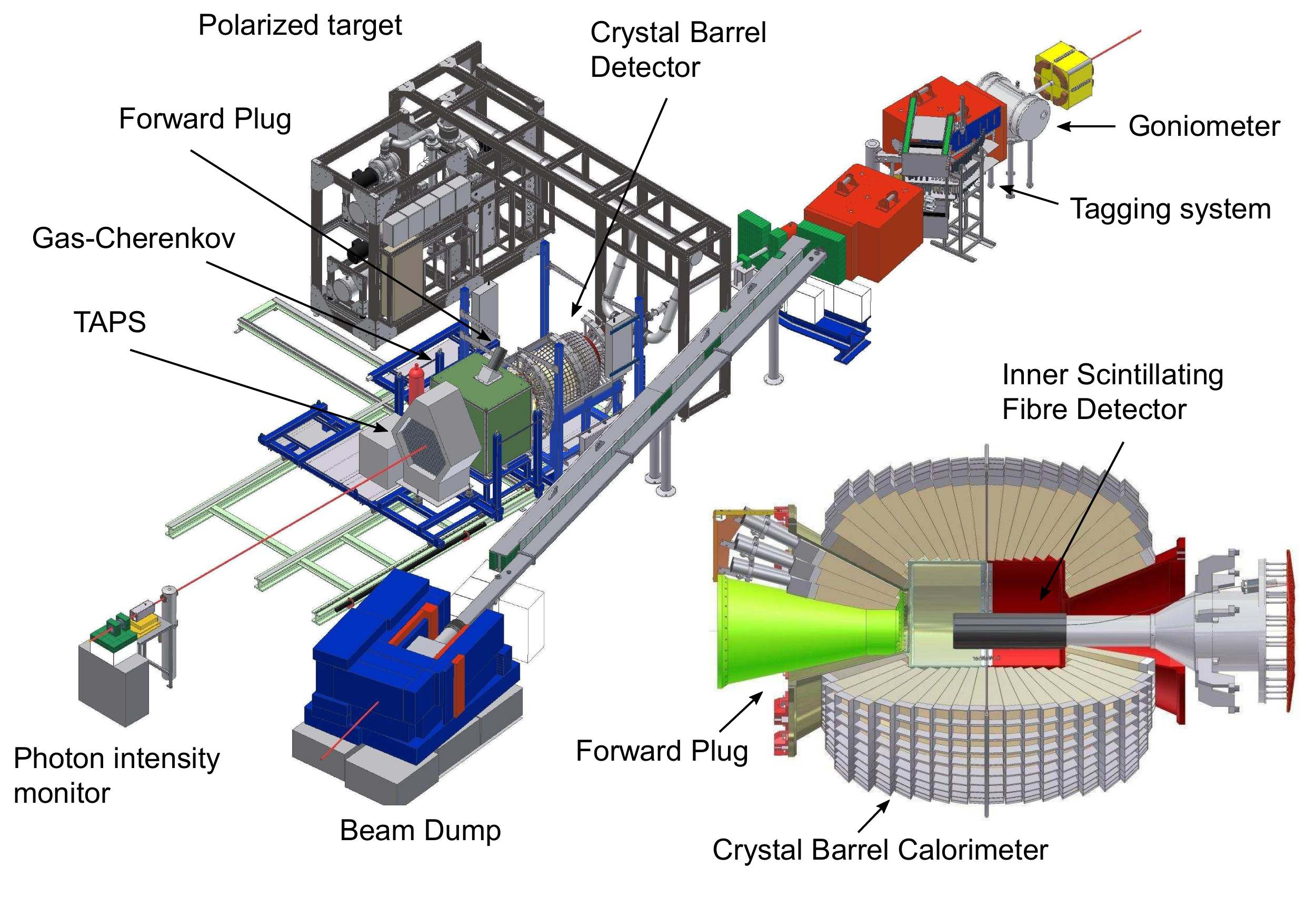}
\caption{\label{fig:setup}Experimental setup of the CBELSA/TAPS experiment.}
\end{center}\end{figure*}
Figure~\ref{fig:setup} shows an overview of the experimental setup.

\subsection{Reconstruction and event selection}
Photon candidates were defined by hits in the calorimeters and no related hit in the scintillation
fiber detector or the plastic scintillation counters. The four-momenta of photons were determined by 
measuring their energies and directions assuming 
that the photons originated from the target center. 
Charged particles were identified by hits in one of the scintillation counters associated with a calorimeter hit.
In the case of the longitudinally polarized target, the electromagnetic background
was considerably lower, and charged particles were also identified by hits in the inner detector only.
In the analysis of data with the transversely polarized target, photon and proton candidates were reconstructed 
from events which had only hits in the calorimeters. Then, the best kinematic combination was chosen
with one meson and one proton in the final state.   

\begin{figure}[h]
\centering%
\includegraphics[width=0.49\textwidth]{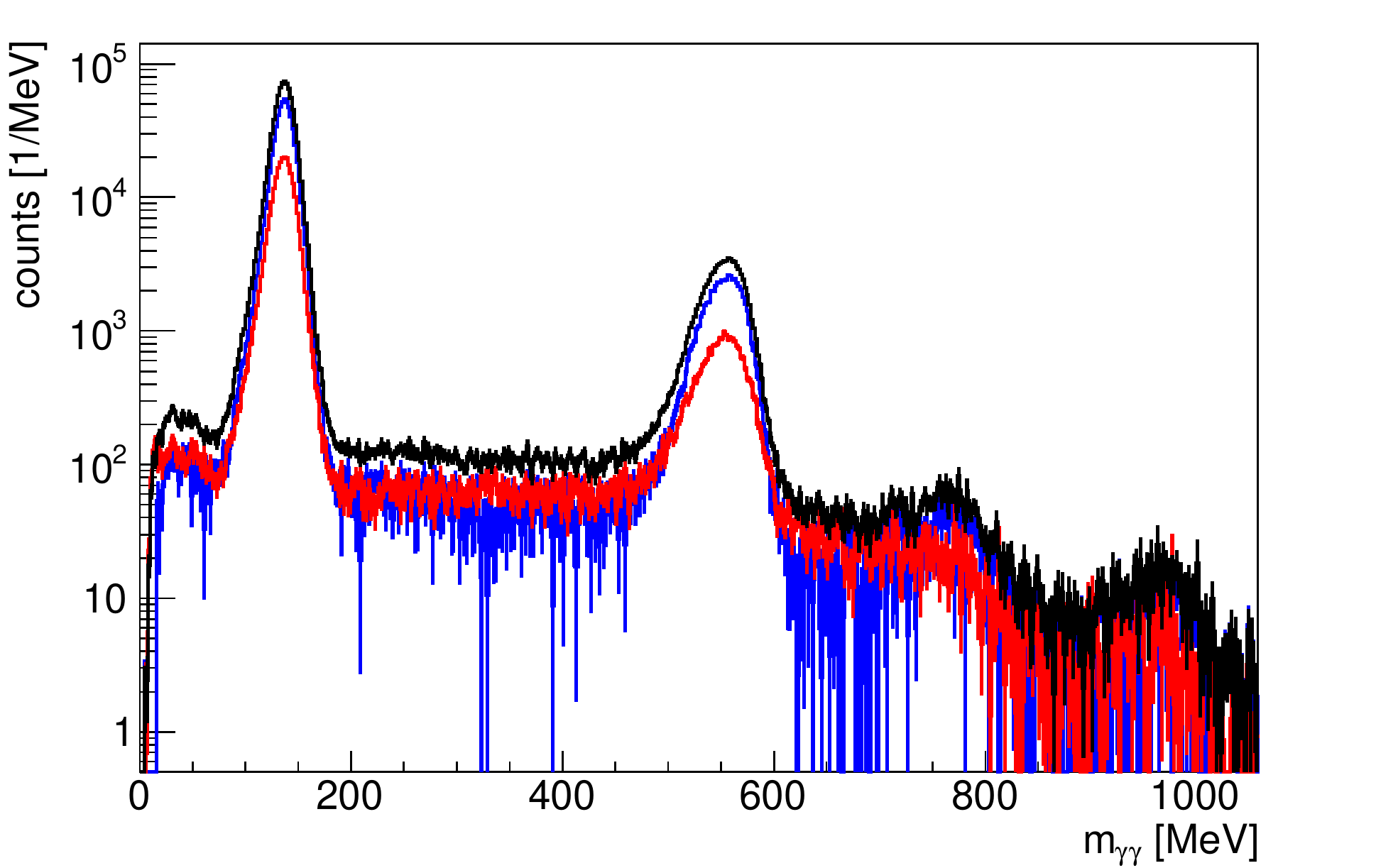}\\[-2ex]
\caption{\label{gg_eta} $\gamma\gamma$-invariant mass distribution for the data with 
transversally polarized target and 
linearly polarized photons, black: butanol data, red: carbon data, blue: difference. 
Random time background already subtracted.}
\end{figure}

Events due to $\gamma p\to \gamma\gamma p$ were selected by choosing 
events satisfying the following criteria: 
two photon and one proton candidates had to be detected;
the invariant mass of the two photons had to agree within {$\pm2\sigma$} with the $\eta$ mass (see Fig.~\ref{gg_eta}); 
the missing mass $X$ from $\gamma p\to \gamma\gamma X$ had to agree with the proton mass within {$\pm2\sigma$}, 
the azimuthal angle between the direction of proton and $\eta$ was requested to be 180$^{\circ}$ within a $\pm2\sigma$ window
(coplanarity), an additional $\pm2\sigma$-cut on the respective polar angle was performed for part of the data sets. 
All these cuts were done taking the energy-dependent width of the respective quantity into accout.  
In addition, a time coincidence was required between the tagger hit and the reaction products, 
and random time background was subtracted.

\subsection{Dilution factor}
\begin{figure}[h]
\centering%
\includegraphics[width=0.242\textwidth]{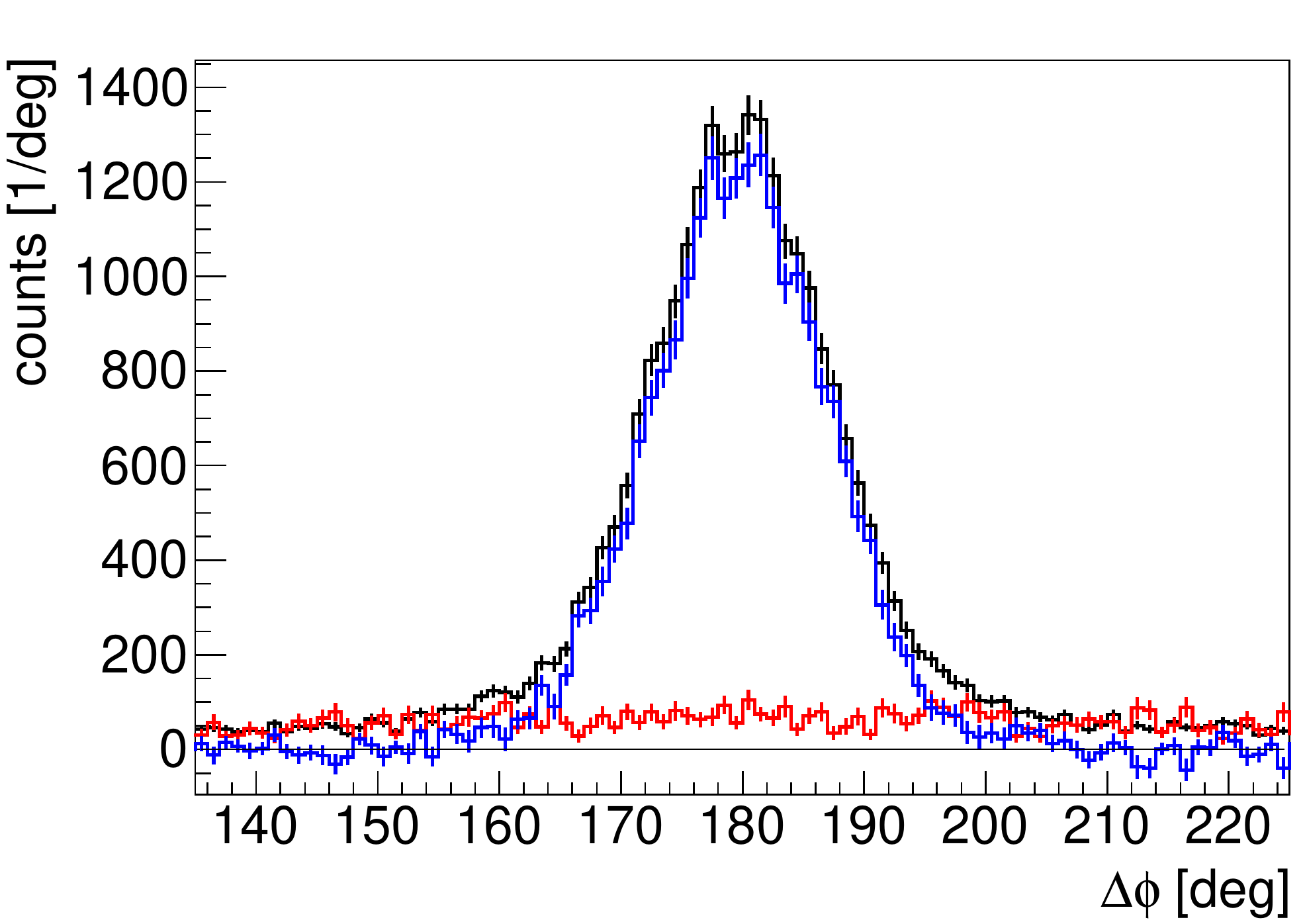}%
\includegraphics[width=0.242\textwidth]{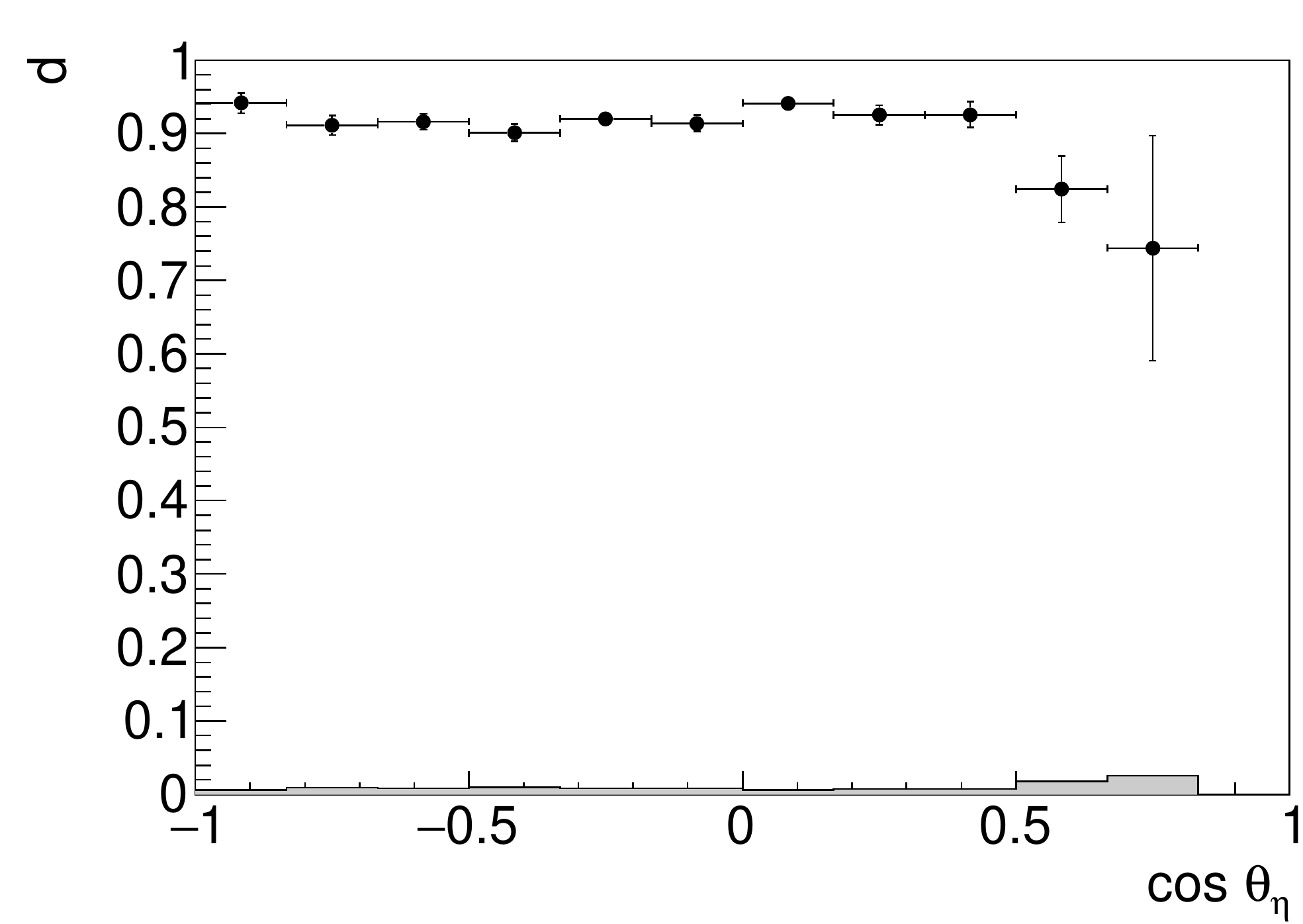}\\[+.5ex]%
\includegraphics[width=0.242\textwidth]{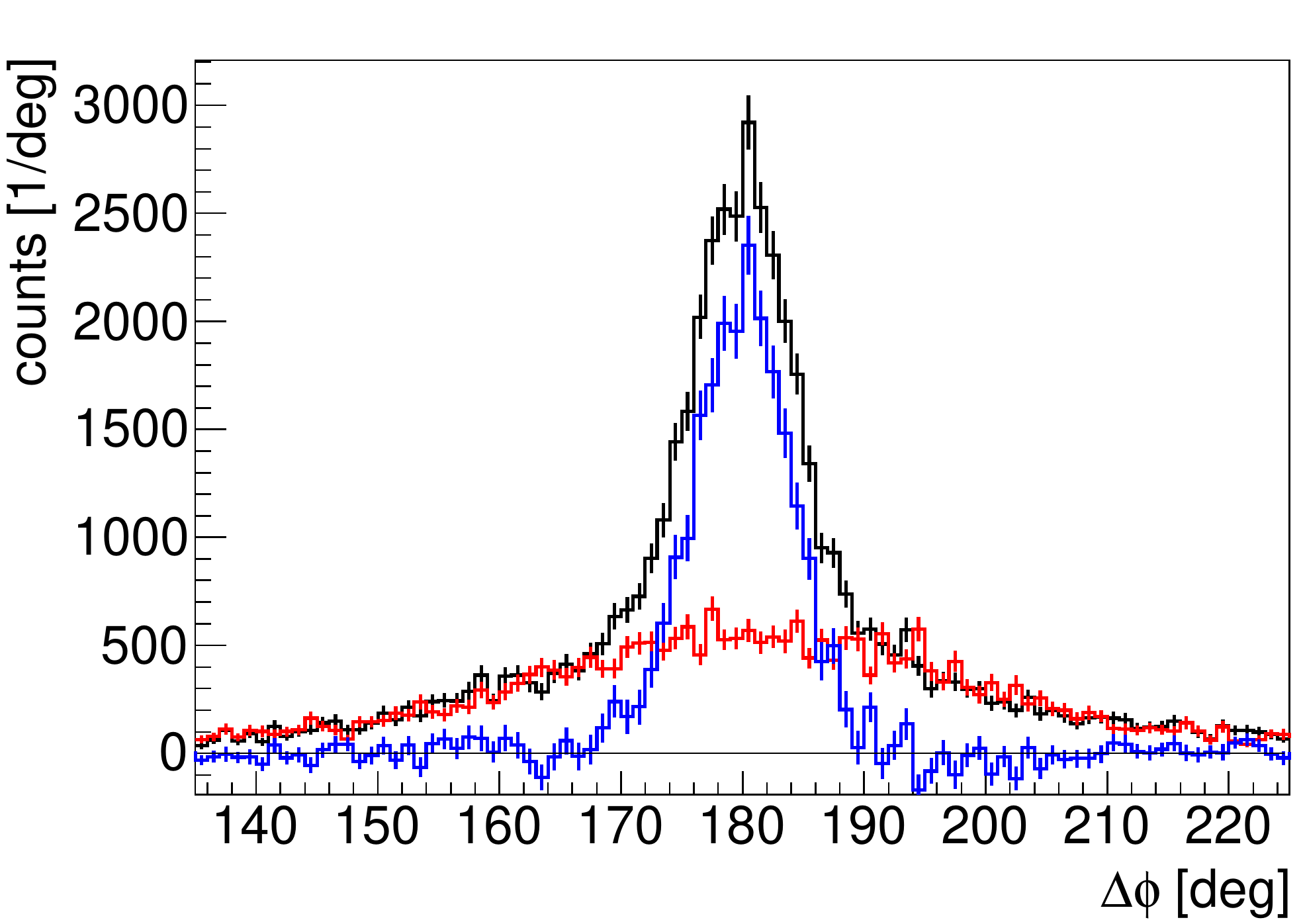}%
\includegraphics[width=0.242\textwidth]{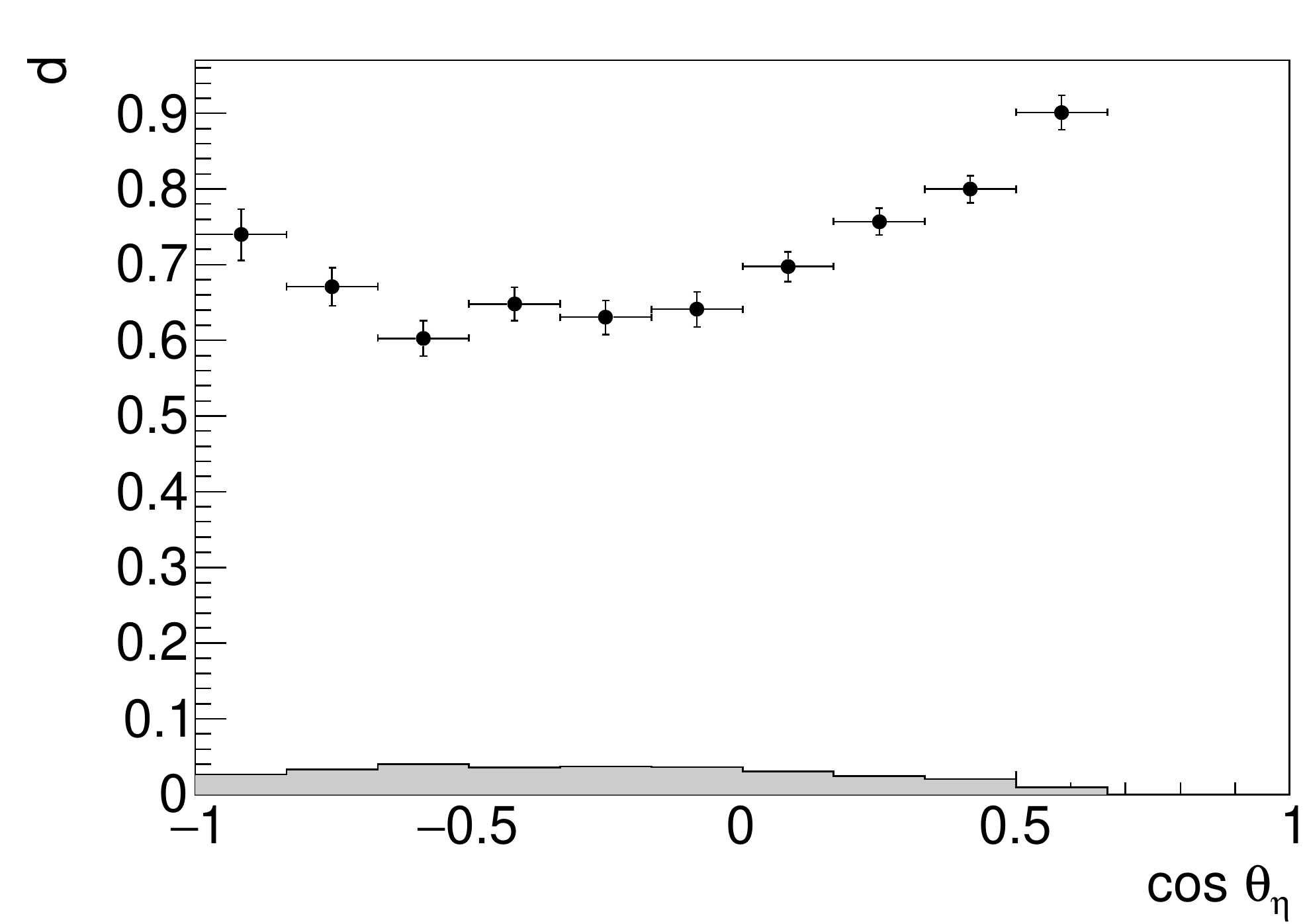}\\[-1ex]
\caption{\label{fig:dil_factor} Left: coplanarity spectra ($T,P,H$-data), 
black: butanol data, red: scaled carbon data, blue: subtracted spectrum (free protons). Right: Dilution factor $d$.
	Upper row: $1513\,\mathrm{MeV} < W < 1531\,\mathrm{MeV}$, lower row: $1660\,\mathrm{MeV} < W < 1716\,\mathrm{MeV}$. The gray bands show the systematic uncertainty due to normalization of carbon data.
}
\end{figure}

In a butanol target, polarizable 
free protons $(f)$ as well as nucleons bound $(b)$ in carbon or
oxygen nuclei contribute to the count rate. The contribution of bound nucleons was 
determined using a carbon foam target within the cryostat with approximately the same 
density as the carbon and oxygen part of the butanol target. The 
coplanarity distribution of events produced off bound nucleons is wider than the one for free
protons. This effect was used to
determine -- for each bin in energy and angle -- the fraction of the reactions off free protons in the 
data collected with the butanol target (see Fig.~\ref{fig:dil_factor}). 
This fraction is called dilution factor 
$$d(E_{\gamma},\cos\theta_{\eta})={N^f}/{(N^f+N^b)}$$
and was determined as $d={(N_{\text{butanol}}-s\cdot N_{\text{carbon}})}/{N_{\text{butanol}}}$.
The carbon normaliztion factor $s$ was determined by comparing the carbon data to the butanol data, excluding kinematic regions where contributions from free protons can be expected.
The dilution factor, as determined for the $T, P, H$-data, is shown in Fig.~\ref{fig:dil_factor}, further examples are given in \cite{Gottschall:2013uha,Gottschall:2019pwo,Hartmann:2014mya,Thiel:2012yj}.

\begin{figure*}[t!]
\begin{center}
\includegraphics[width=0.999\textwidth]{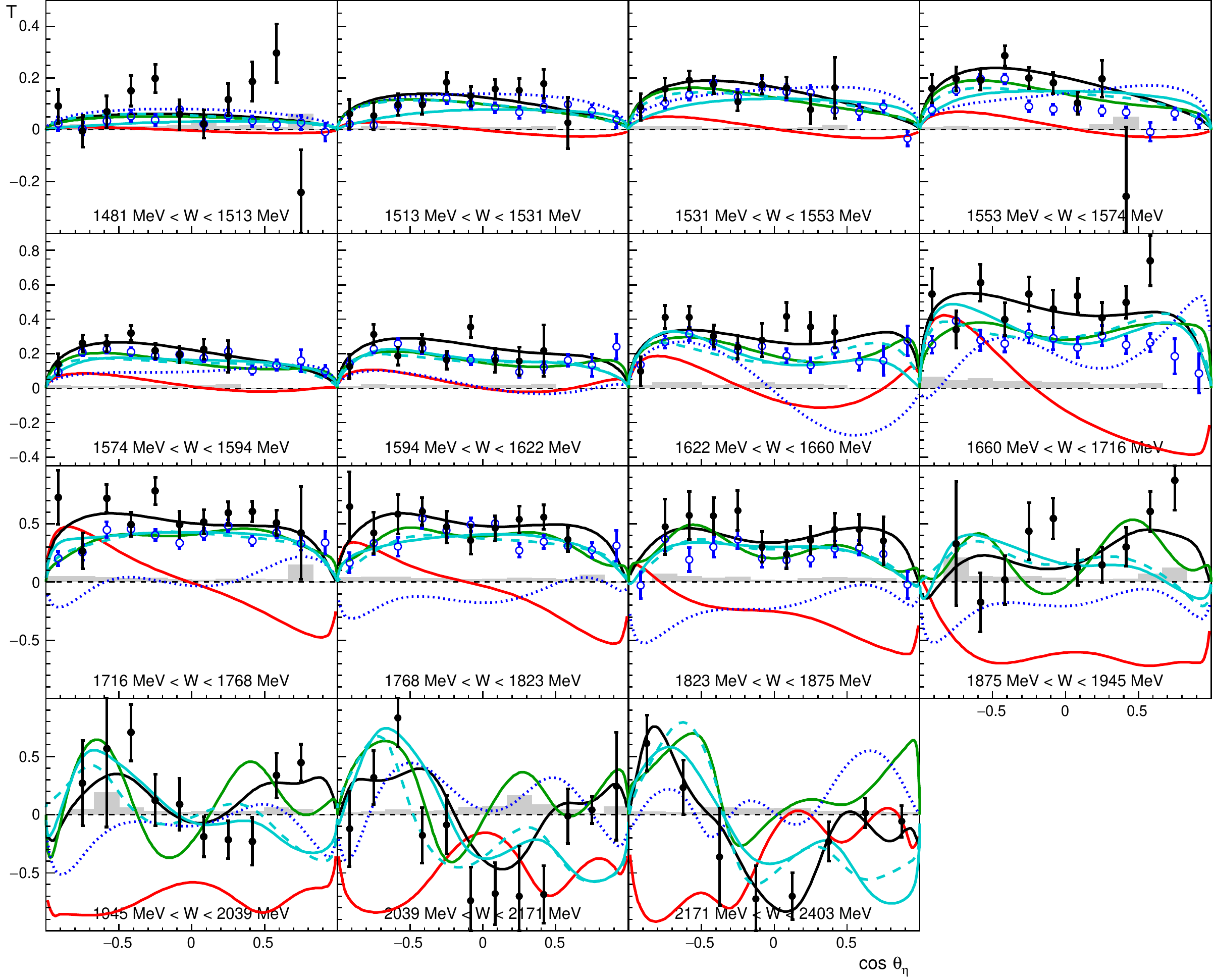}\vspace{-6mm}
\end{center}
\caption{\label{fig:data_T}The polarization observable $T$ as function of $\cos\theta_\eta$, where $\theta_\eta$ is the $\eta$ production angle in the cms for different cms energy ranges. The curves represent different models. Black: BnGa refit; red: BnGa2011-02 \cite{Anisovich:2011fc}; green: MAID2018~\cite{MAID2018}; dark blue (dotted): SAID (GE09)~\cite{SAID-GE09}; light blue: J\"uBo\,2015~\cite{deborah} (dashed) and J\"uBo\,2015-3~\cite{Senderovich:2015lek} (solid). The different PWA curves are calculated at the central energy of each bin.  
(J\"uBo\,2015-3, included the recent CLAS-data on $E$~\cite{Senderovich:2015lek},  
MAID2018~\cite{MAID2018} in addition to $E$ from CLAS~\cite{Senderovich:2015lek}, also 
$T$ and $F$ from MAMI~\cite{Akondi:2014ttg}.)
The systematic errors due to photon and proton polarization, dilution factor, and background contamination are shown as a gray band. 
Recent data from MAMI~\cite{Akondi:2014ttg} are shown for comparison as blue open points (due to different binning, the energies differ by up to $\Delta W = 14$~MeV).}
\end{figure*}

\begin{figure*}[t!]
\begin{center}
\includegraphics[width=0.975\textwidth]{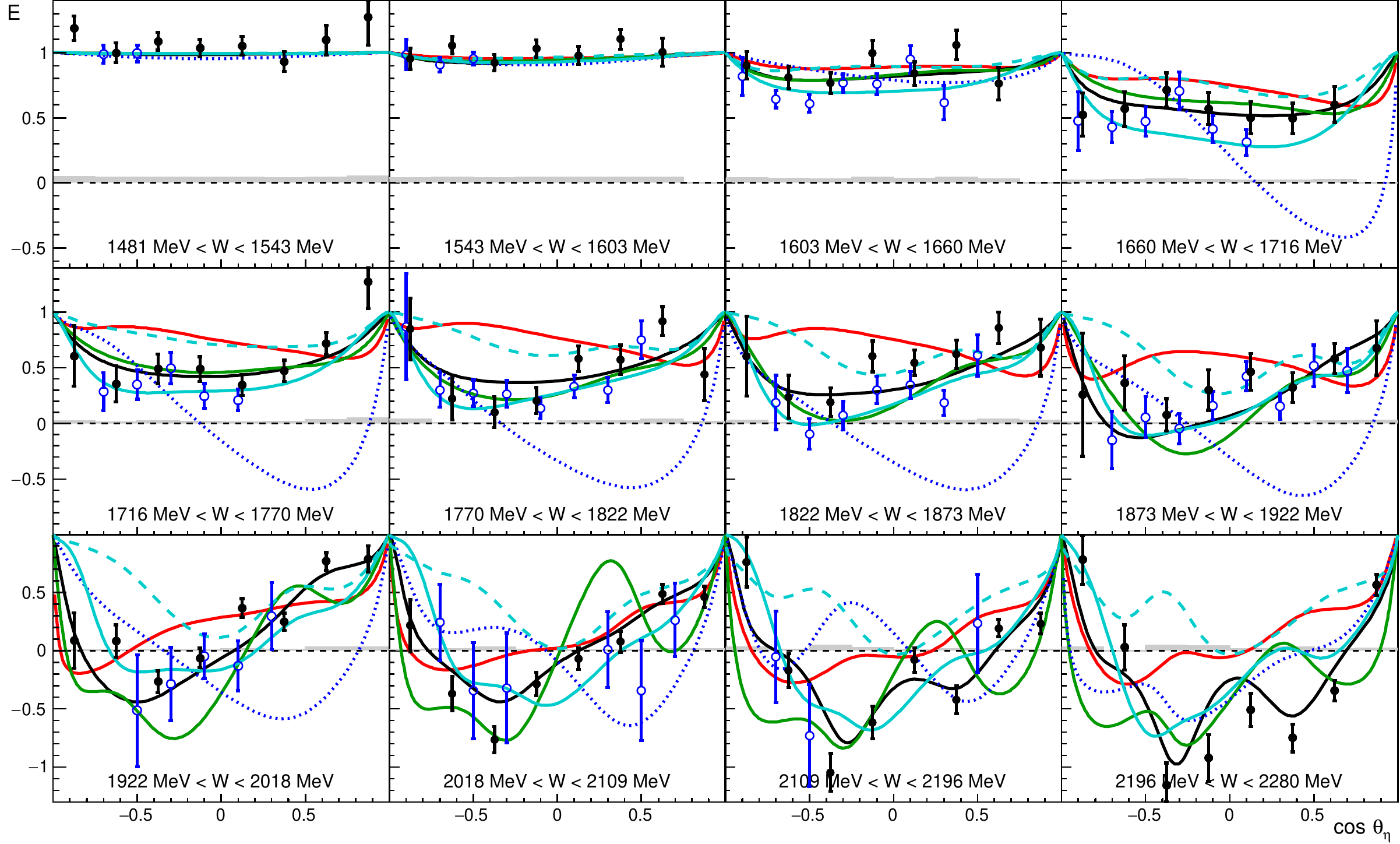}\vspace{-6mm}
\end{center}
\caption{\label{fig:data_E}The double polarization observable $E$ as function of $\cos\theta_\eta$, where $\theta_\eta$ is the $\eta$ production angle in the cms for different cms energy ranges. See caption of 
Fig.~\ref{fig:data_T} for an explanantion of the symbols. Recent data from CLAS~\cite{Senderovich:2015lek} are shown for comparison as blue open points (due to different binning, the energies differ by up to half of the bin size).}
\begin{center}\vspace{+4mm}
\includegraphics[width=0.47\textwidth]{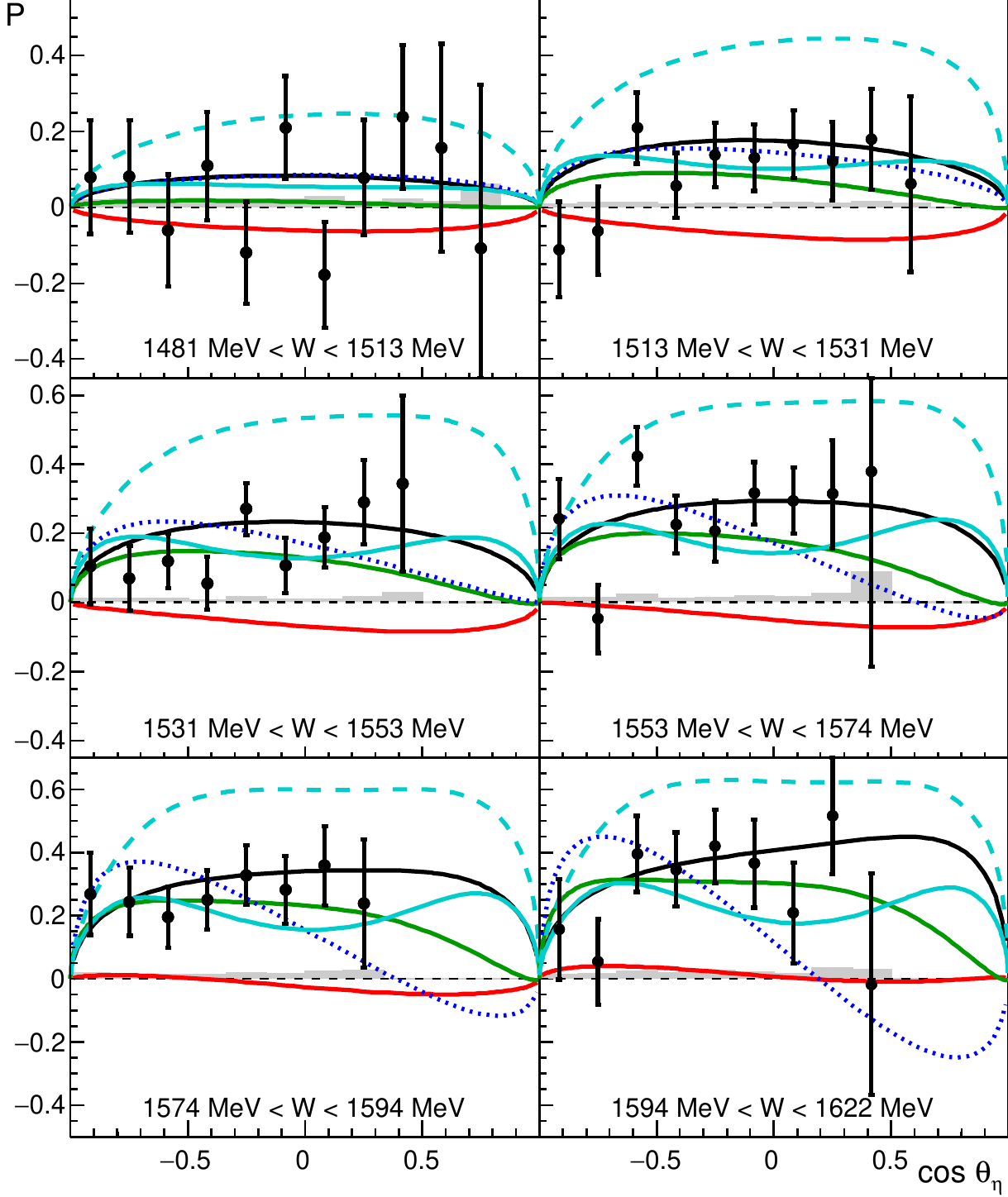} \hspace{1.5em}
\includegraphics[width=0.47\textwidth]{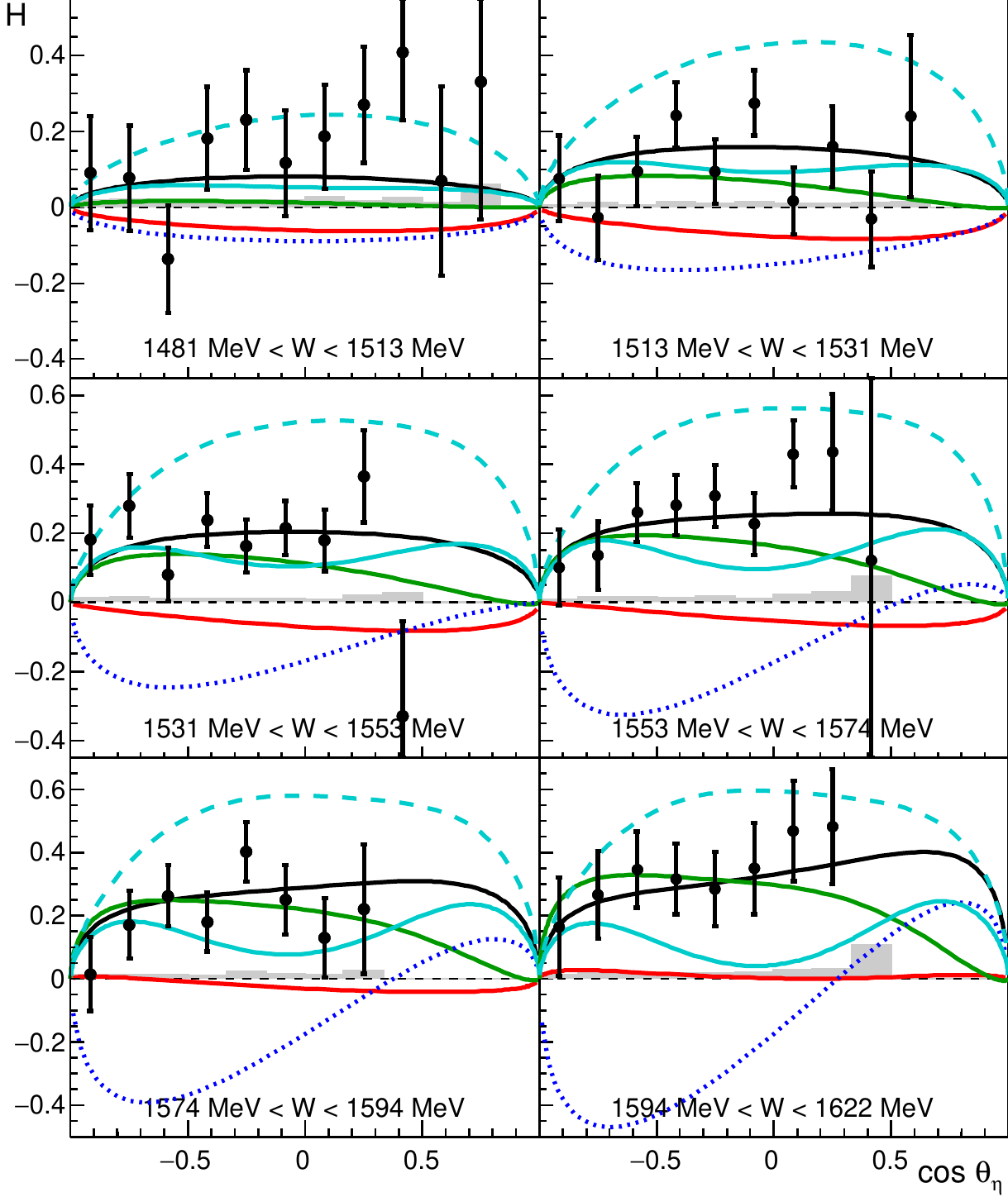}
\vspace{-6mm}
\end{center}
\caption{\label{fig:data_PH}The double polarization observables $P$ and $H$ as functions of $\cos\theta_\eta$, where $\theta_\eta$ is the $\eta$ production angle in the cms for different cms energy ranges. See caption of 
Fig.~\ref{fig:data_T} for an explanantion of the symbols.}
\end{figure*}

\subsection{Polarization observables}
The helicity asymmetry $E$ requires circularly polarized photons and longitudinally
polarized protons. It can be determined as  
\begin{equation} {E}=
\frac{N_{1/2}-N_{3/2}}{N_{1/2}+N_{3/2}}\cdot \frac{1}{d}\cdot
\frac{1}{p_{\odot}p_T}\,,
\label{counts}
\end{equation}
where $N_{1/2}$ and $N_{3/2}$ are the 
number of events observed with photon and target polarization in opposite or parallel directions, 
normalized to the 
corresponding number of incident photons.

$G$ can be deduced from the correlation between the photon polarization plane and the scattering plane 
for protons polarized along the direction of the incoming photon.
The number of events $N$ 
as a function of the azimuthal angle $\phi$ between the two planes 
is given by
\begin{eqnarray}
\hspace{-3mm}\frac{N(\phi)}{N_{0}} =1- p_\gamma \cdot\left[
\varSigma_{\rm eff} \cos(2\phi) -d\,p_T G 
\sin(2\phi)\right],
\label{pol-h}
\end{eqnarray}
where $N_0$ is given by averaging $N(\phi)$ over $\phi$. $\varSigma_{\rm eff}$ mixes 
the beam asymmetry of free and bound nucleons.

The observables $T$, $P$, and $H$ can be measured simultaneously when a transversely polarized 
target and a linearly polarized photon beam are used. In that case, the azimuthal distribution of events is given by
\begin{align}
\frac{N(\phi)}{N_{0}} = 1 &- p_{\gamma}\varSigma_{\rm eff}\cos(2\phi) + d\,p_{T} T \sin(\phi-\alpha) \nonumber \\[-1.5ex]
& - d\,p_{T}\,p_{\gamma} P\cos(2\phi)\sin(\phi-\alpha) \nonumber \\
& + d\,p_{T}\,p_{\gamma} H\sin(2\phi)\cos(\phi-\alpha),
\end{align}
where $\alpha$ is the azimuthal angle between the target polarization vector and the photon polarization plane. 
$T$, $P$, and $H$ are determined, for each $(E_\gamma,\cos\theta_\eta)$ bin, 
from an event-based maximum likelihood fit~\cite{tobias-jan_in_prep} to the measured azimuthal distribution of events.

\subsection{Systematic Uncertainties}
The data-taking periods, the target and beam polarization as well as the analyses methods used to extract 
the different observables were not identical for the data presented here; therefore the systematic uncertainties 
for the different data sets are discussed separately in the following. 
The systematic uncertainties of all observables include contributions from possible background events,
the determination of the dilution factor, and the polarization degrees of target (all observables) and beam (only $E,G,H,P$).

The polarization of the circularly polarized photon beam was calculated using the polarization transfer from the longitudinally polarized incident electron beam \cite{olsen_maximon}. The electron polarization was measured in parallel to data-taking using a M\o{}ller polarimeter with a relative systematic uncertainty of 3.3\% \cite{Gottschall:2019pwo}. 
The polarization of the linearly polarized photon beam was determined from the measured photon intensity spectrum using a software \cite{Elsner:2008sn} based on the analytic bremsstrahlung (ANB) calculation \cite{anb}. For the measurement of $G$ the relative uncertainty was 5\%. For the observables $P$ and $H$, measured only up to $E_\gamma = 933$\,MeV, a relative uncertainty of 4\% was achieved. 

The polarization of the dynamically polarized target protons was measured using an NMR system \cite{reicherz_nmr}.
It was calibrated using the proton polarization in thermal equilibrium. 
A relative systematic uncertainty of 2\% was reached for all data sets.

The determination of the dilution factor $d$ relies on the relative 
normalization of the carbon data. A conservative uncertainty of 10\% 
was assumed for the normalization factor $s$. Close to threshold, where 
$d > 0.9$, this yields a systematic uncertainty $\Delta d < 0.01$. Since 
$d$ decreases with energy, its uncertainty increases up to $\Delta d = 0.05$ 
for $E_\gamma > 2$\,GeV. 
A systematic uncertainty of comparable magnitude was determined 
for the observable $G$ using a different method. Here either 
carbon or carbon and LH$_2$-data were used in combination 
with the butanol data to determine $d$. The resulting 
differences were considered as systematic uncertainties. 

Background contamination of the event samples was found to be below $2\%$ in most bins. Only a few bins at the edge of the detector acceptance exhibit more background, up to 5\%--15\%, depending on the data set and exact selection criteria used.
For the observable $E$ the background was found to be unpolarized, and the values of $E$ were corrected accordingly. For the other observables, the asymmetry of the background could not be constrained significantly because of the limited size of the event samples and the small background contribution. Instead, the relative background contamination was taken as an additional systematic uncertainty of the observables.

\section{Results {\color{magenta} }}
\subsection{Observables}
\begin{figure}[pt]
\begin{center}
\hspace{-1mm}
\includegraphics[width=0.47\textwidth]{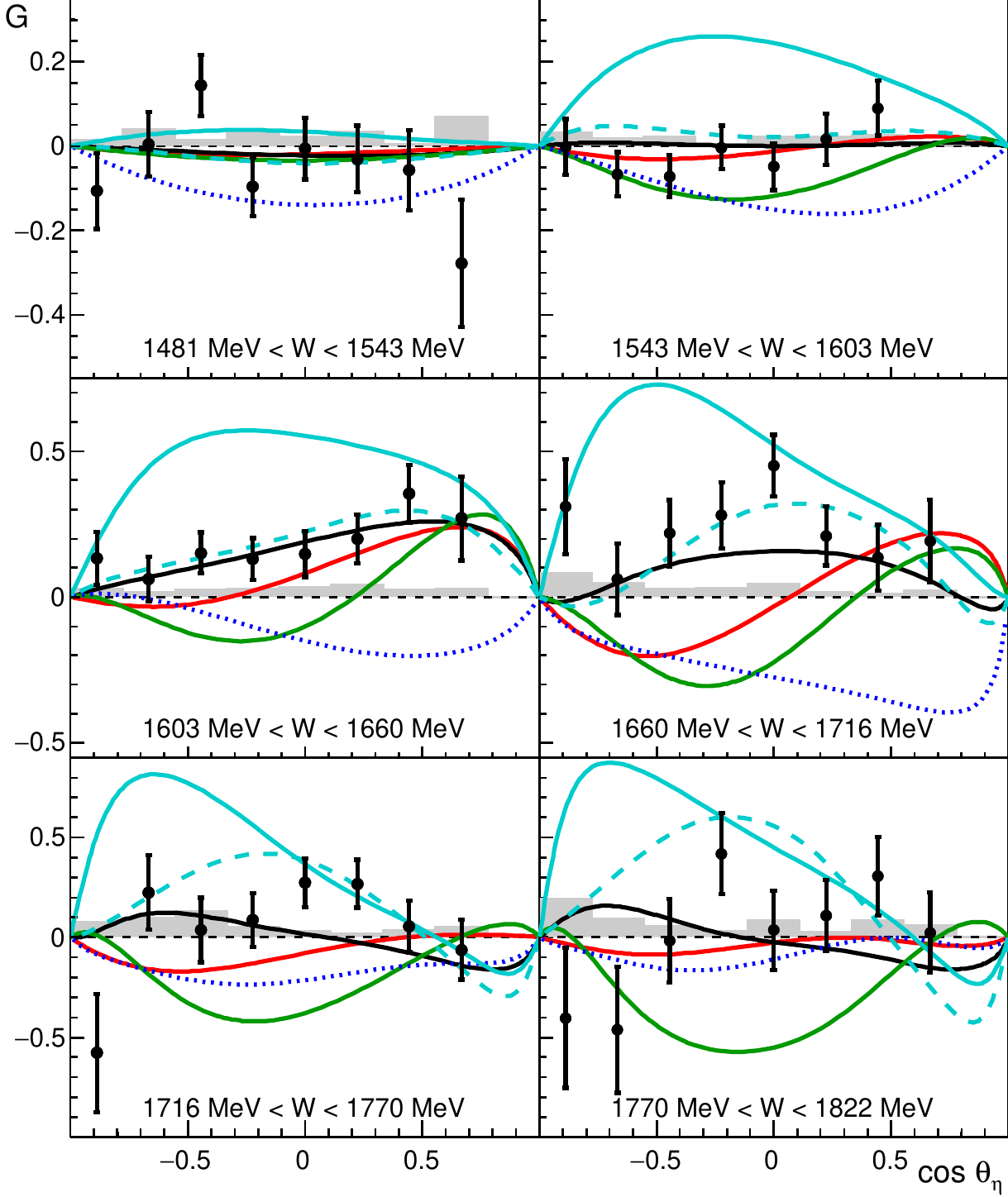}
\vspace{-6mm}
\end{center}
\caption{\label{fig:data_G}The double polarization observable $G$ as function of $\cos\theta_\eta$, where $\theta_\eta$ is the $\eta$ production angle in the cms for different cms energy ranges. See caption of 
Fig.~\ref{fig:data_T} for an explanantion of the symbols.}
\end{figure}

Figures~\ref{fig:data_T}--\ref{fig:data_G} show the resulting double polarization observables $T$, $E$, $P$, $H$, and $G$. 
Only for $E$ and $T$, earlier data exist that cover more than 
a few energy and angular bins. 
CLAS has published data on $E$~\cite{Senderovich:2015lek}. Our data extend 
the energy and angular range of the CLAS data. Within uncertainties, the CLAS 
data are in good agreement with our findings (see Fig.~\ref{fig:data_E}). 

In the low-energy region, $E$ is expected to be close to $+1$, since this region 
is dominated by one single resonance with spin-parity $J^P=1/2^-$. The 
asymmetry should not exceed one. As visible in Fig.~\ref{fig:data_E}, three of the 
data points in the first energy bin exceed one beyond their statistical 1$\sigma$ error. 
Averaging all data points in the first energy bin yields $E = 1.05\pm 0.03_\mathrm{stat} \pm 0.04_\mathrm{sys}$, 
assuming full correlation of the systematic uncertainties. Adding both 
uncertainties in quadrature results in a deviation from 1 by $1\sigma$. In later 
fits, tests have been performed rescaling the data so that the error weighted mean of the 
data points  in the first bin is one. The changes observed in the fit results are covered by the errors given.   


Data on $T$~\cite{Akondi:2014ttg} are available from MAMI. 
Our data extend the energy range of the MAMI data. However,
the comparison of our $T$-data  and the MAMI  $T$-data reveals serious 
discrepancies (see Fig.~\ref{fig:data_T}). On average, the MAMI $T$-asymmetry are smaller than our results 
by a factor 0.7. If an overall scaling 
factor of 0.7 is introduced, the two data sets agree nicely. 
The difference $\Delta T= T_{\rm MAMI}/0.7 - T_{\rm ELSA}$, normalized to the statistical error, results 
in a Gaussian distribution centered at zero with a width of $\sigma=1.09\pm0.07$. 
The origin of this discrepancy is not understood. The same analysis based 
on the same CBELSA/TAPS-data set was used to derive $T$ for 
$\gamma p \to p \pi^0$~\cite{Hartmann:2014mya,Hartmann:2015kpa}. 
These results and the MAMI results for $T(\gamma p \to p \pi^0)$~\cite{Annand:2016ppc} are fully 
consistent.  

$T$, $P$, $H$, and $G$ are small as 
expected and as predicted by most partial-wave analyses. 
The data sets are shown in comparison to various PWA-predictions. 
Already at low energies, below $E_\gamma$=1\,GeV ($W$=1.660\,GeV), the PWA-predictions 
show significant deviations from the data, above $E_\gamma$=1\,GeV, the data and the predictions 
diverge: none of the partial-wave analyses predicted all 
observables with reasonable accuracy. 
Large deviations from the data are observed for 
the predictions from 
SAID~\cite{SAID-GE09}, BnGa2011~\cite{Anisovich:2011fc},  
and the J\"uBo model~\cite{deborah} (J\"uBo\,2015-3, includes the 
recent CLAS-data on $E$~\cite{Senderovich:2015lek}). MAID2018~\cite{MAID2018}, 
which includes recent data on $E$ \cite{Senderovich:2015lek} and $T$, $F$~\cite{Akondi:2014ttg}, 
exhibits fewer deviations but still fails to predict e.g. $G$.
The comparison shows how important these new data are 
to constrain the amplitudes for photoproduction of $\eta$ mesons off protons. 

\subsection{PWA fits }
These data, and further new $\gamma p \to p\eta$-data 
from MAMI $\frac{d\sigma}{d\Omega}$~\cite{Kashevarov:2017kqb},  ($T, F$)~\cite{Akondi:2014ttg},  
CLAS $\Sigma$~\cite{Collins:2017sgu}, $E$~\cite{Senderovich:2015lek}, and 
CBELSA/ TAPS $\Sigma$~\cite{Farah-2019} as well as the $\eta^\prime$-data sets used 
in~\cite{Anisovich:2017pox} are included in the data base used in~\cite{Anisovich:2011fc}.  
The full data base 
also includes 
the GWU $\pi N$ partial-wave amplitudes and data on the pion and photo-produced 
$\pi N$, $\eta N$, and $K Y$ (Y:\,hyperon) final states \cite{Anisovich:2004zz,Anisovich:2006bc,Anisovich:2007zz}. 
Also the data on $\pi^-p\to\eta n$ \cite{Richards:1970cy,Brown:1979ii,Prakhov:2005qb} are included.  (For a complete list 
of data sets included into the BnGa-PWA, see also:~\cite{pwa-webseite}.)
In the study presented here, the couplings to the $\pi^0\pi^0 N$
and $\pi^0\eta p$ final state are frozen to values derived in~\cite{Sokhoyan:2015fra,Gutz:2014wit}.
For the differential cross sections scaling factors were used in the fit to take care about 
normalization inconsistencies. 

\subsection{$N^*\to N\eta$ decays {\color{magenta} }}
Table~\ref{table:br} presents the resonances used in the fit and the 
resulting branching ratios for $N^*\to N\eta$ decays.
The uncertainties result from a variation of the fit hypothesis, in particular
the inclusion of additional resonances or the exclusion of minor resonances 
in the fit.
We studied the effect the systematic difference visible in $T$ between 
the MAMI ($T,F$) and the CBELSA/TAPS data ($T,P,H$) might have on the fit. 
To do so we allowed either for scaling factors in the MAMI $T,F$-data or 
in the CBELSA/TAPS $T,P,H$-data. When scaling factors for the 
MAMI-data were admitted, they optimized at 0.73 for $T$ and 0.67 for $F$ resulting 
in a $\chi^2/N_{\rm data}$ for the two data sets of 1.608 and 1.464. These $\chi^2/N_{\rm data}$ 
values are almost identical to the values of the final fit ($\chi^2/N_{\rm data}$=1.609, 1.465, respectively) 
where we fixed the scaling factor to 0.7 as determined experimentally for $T$. 
Next, the MAMI $T,F$-data were included in the fit without scaling factors, while scaling 
factors were introduced for the CBELSA/TAPS $T,P,H$-data. The $\chi^2/N_{\rm data}$ for the MAMI $T,F$-data 
was found to significantly worse ($\chi^2/N_{\rm data}$=3.3, 2.5, respectively), the $\chi^2/N_{\rm data}$ for the 
CBELSA/TAPS $T$ got slightly worse, it improved slightly for $P$ and remained the same for $H$.
At the same time the overall weigthed~\cite{Anisovich:2011fc} $\chi^2/N_{\rm data}$ of the fit for all $\gamma p \to p \eta$-data sets 
increased from $\chi^2/N_{\rm data}$=1.42 (final fit) to $\chi^2/N_{\rm data}$=1.5. 
Obviously, the fit constrained by the existing $p\eta$-data sets finds a better consistency 
between the different data sets when  MAMI ($T,F$)-data were scaled. 
The variations in the branching ratios (BR) and helicity 
amplitudes ($A_{1/2}$, $A_{3/2}$) found in these studies 
are covered by the errors given in Tab.~\ref{table:br}. 

\begin{table}[b!]
\renewcommand{\arraystretch}{1.4}
\caption{\label{table:br}
Branching ratios (BR) for $N^*\to N\eta$ decays and the photon helicity amplitudes
$A_{1/2}$, $A_{3/2}$ of nucleon resonances, both calculated at their pole positions. 
The helicity amplitudes are given in units of GeV$^{-1/2}$.
Small numbers below the BRs give the
RPP\,2017~\cite{PDG2017} (representing the status before new $\eta$ (double) polarization data became available from CLAS, MAMI and CBELSA/TAPS). $A_{1/2}$, $A_{3/2}$ at the pole positions are complex numbers. Here we give the absolute value with a positive sign if the phase falls between $-45^\circ$ and $+45^\circ$, a negative sign for $135^\circ<\phi<225^\circ$ and ``*'' 
otherwise. 
 The small values below the helicity amplitudes 
$A_{1/2}$, $A_{3/2}$ give values from~\cite{Sokhoyan:2015fra}, if available, otherwise marked by 
(R) values from~\cite{Ronchen:2014cna} are given, 
since no PDG-estimates exist for this quantity. 
\vspace{-2mm}
}
{\footnotesize
\begin{center}
\begin{tabular}{|cc|cc|}
\hline\hline
Res. \,   & \hspace{-2mm}BR{\scriptsize($N^*\to N\eta$)}\hspace{-2mm}\, & Res. \, & \hspace{-2mm}BR{\scriptsize($N^*\to N\eta$)}\hspace{-2mm}\, \\[-1ex]
$A_{1/2}$ & $A_{3/2}$     & $A_{1/2}$ & $A_{3/2}$\\
\hline
$N(1535)$                 &0.41\er0.04      &
$N(2120)$                 &$\leq$0.01       \\[-1.5ex]
\tiny$1/2^-$              &\tiny 0.32-0.52  &
\tiny$3/2^-$              & -               \\[-1ex]
+0.096\er0.008             & -               &
+0.110\er0.045             &+0.130\er0.050    \\[-1.5ex]
\tiny +0.114\er0.008       & -               &
\tiny +0.130\er0.045       &\tiny +0.160\er0.060  \\\hline
$N(1650)$                 &0.33\er0.04       &
$N(1720)$                 &0.03\er0.02         \\[-1.5ex]
\tiny$1/2^-$              &\tiny 0.14 - 0.22 &
\tiny$3/2^+$              &\tiny0.01-0.05\\[-1ex]
+0.032\er0.006             & -                &
+0.090\er0.035             &*0.120\er0.035     \\[-1.5ex]
\tiny +0.032\er0.06 &-                 &
\tiny +0.115\er0.045       &\tiny *0.140\er0.040\\\hline
$N(1895)$                 &0.10\er0.05       &
$N(1900)$                 &0.02\er 0.02       \\[-1.5ex]
\tiny$1/2^-$              &\tiny 0.15-0.27 &
\tiny$3/2^+$              &\tiny 0.02-0.14    \\[-1ex]
-0.030\er0.010            & -                 &
*0.026\er0.014             &*0.090\er0.020     \\[-1.5ex]
\tiny -0.015\er0.006      & -                 &
\tiny *0.026\er0.014       &\tiny *0.070\er0.030\\\hline 
$N(1710)$                 &0.18\er0.10        &
$N(1675)$                 &0.005\er0.005      \\[-1.5ex]
\tiny$1/2^+$              &\tiny 0.10 - 0.50  &
\tiny$5/2^-$              &\tiny $<$0.01\\[-1ex]
*0.035\er0.015             & -                 &
+0.020\er0.004             &  { +0.028\er0.005} \\[-1.5ex]
\tiny *0.028$^{+0.009}_{-0.002}$(R)
& -                 &
\tiny +0.022\er0.003       &\tiny +0.028\er0.006 \\\hline
$N(1880)$                 &0.18\er0.08        &
$N(2060)$                 &0.06\er0.02     \\[-1.5ex]
\tiny$1/2^+$              &\tiny 0.05 - 0.55&
\tiny$5/2^-$              &\tiny 0.02 - 0.06\\[-1ex]
+0.040\er0.015             &-                  &
+0.070\er0.010             &+0.070\er0.020      \\[-1.5ex]
\tiny -                   &-                  &
\tiny+0.064\er0.010        & \tiny+0.060\er0.020\\\hline
$N(2100)$                 &0.30\er0.15        &
$N(1680)$                 &0.002\er0.001   \\[-1.5ex]
\tiny$1/2^+$              & \tiny seen                &
\tiny$5/2^+$              & \tiny $<$0.01  \\[-1ex]
*0.010\er0.004            & -               &
-0.014\er0.002            & +0.134\er 0.005 \\[-1.5ex]
\tiny *0.011\er0.004       & -               &
\tiny-0.013\er0.003       &\tiny +0.135\er 0.005 \\\hline
$N(1520)$                 & $<0.001$        &
$N(2000)$                 &0.02\er0.02    \\[-1.5ex]
\tiny$3/2^-$              & \tiny $<$0.01  &
\tiny$5/2^+$              & \tiny $<$0.04 \\[-1ex]
-0.024\er0.004            & +0.128\er 0.006   &
+0.015\er0.006             &-0.043\er0.008    \\[-1.5ex]
\tiny-0.023\er0.004       &\tiny+0.131\er 0.006&
\tiny+0.033\er0.010       & \tiny-0.045\er0.008\\\hline
$N(1700)$                 &0.01\er 0.01       &
$N(2190)$                 &0.04\er0.02      \\[-1.5ex]
\tiny$3/2^-$              & \tiny seen                 &
\tiny$7/2^-$              &   \tiny seen              \\[-1ex]
*0.045\er0.012             &-0.055\er0.012     &
-0.070\er0.020            &+0.039\er0.007      \\[-1.5ex]
\tiny *0.047\er0.016 &\tiny-0.041\er0.014&
\tiny -0.068\er0.005      &\tiny +0.025\er0.010\\\hline
$N(1875)$                 & 0.10\er0.06  &
$N(1990)$                 &0.01  \er 0.01      \\[-1.5ex]
\tiny$3/2^-$              &\tiny $<$0.01 &
\tiny$7/2^+$              &  -               \\[-1ex]
*0.008\er0.006             & *0.008\er0.004    &
+0.070\er0.020             &+0.044\er0.008     \\[-1.5ex]
\tiny *0.017\er0.009       &\tiny-0.008\er0.004&
\tiny *$0.010^{+0.011}_{-0.006}$(R) &\tiny+$0.053^{+0.023}_{-0.028}$(R)\\\hline\hline
\end{tabular}
\end{center}}\vspace{4mm}\vspace{-9mm}
\renewcommand{\arraystretch}{1.0}
\normalsize\noindent
\end{table}
There are a few remarkable observations: 
The $N\eta$-BR for $N(1535)1/2^-$ is now $0.41\pm0.04$ (instead of the most
recent PDG value of $0.42^{+0.13}_{-0.12}$). Since the statistical error is negligible compared to
the systematic error, we quote only the latter one. Note that our error includes the 
uncertainty due to the $N(1535)1/2^-$ helicity amplitude
$A_{1/2}=(0.096\pm0.008)$ GeV$^{-1/2}$. Second, there is a significant change in  
the $N(1650)1/2^-\to N\eta$ branching ratio: it changes from $0.05$--$0.15$~(RPP\,2014) and 
$0.14$--$0.22$~(RPP\,2017) to $0.33\pm0.04$ in our present fit, reducing substantially
the puzzling difference in the magnitude of the $N\eta$ branching ratios of $N(1535)1/2^-$ and $N(1650)1/2^-$. 
Furthermore, also the $N(1900)3/2^+\to N\eta$ branching ratio changed from $\approx 0.12$~(RPP\,2014) 
and $0.02$--$0.14$ (RPP\,2017) to $0.02\pm0.02$. The $N(1875)3/2^-\to N\eta$ branching ratio is now found to be $0.10\pm0.06$. 
All other values are well within the earlier error bars; some
$N^*{\to}N\eta$ branching ratios are new (even though rather small). 
The $N(1710)1/2^+\to N\eta$  branching ratio settles at $0.18\pm0.10$, well inside 
its previous range $0.05$--$0.55$, while $N(1720)3/2^+$
contributes very little. These results clearly show the power of polarization observables to constrain PWAs;
an earlier PWA~\cite{Crede:2003ax} not including these indicated a large $N(1720)3/2^+$ contribution.

\subsection{The $N(1650)1/2^-\to N\eta$ branching ratio {\color{magenta} }}
The large change in the $N(1650)1/2^-\to N\eta$ branching ratio deserves a more detailed 
discussion. Here we restrict ourselves to a discussion of analyses that include data
on photoproduction. The data on reaction $\pi^-p\to\eta n$ are not sufficiently precise 
to allow for an unambiguous separation of the contributing partial waves. 

The BnGa group reported a $N\eta$ branching ratio of $\BR(N\eta)=0.18\pm0.04$ 
and a helicity coupling of $A_{1/2}=0.033\pm0.007\,$GeV$^{-1/2}$~\cite{Anisovich:2011fc}. 
The helicity coupling and its error were estimated from 12 fits with acceptable 
$\chi^2$, which made different assumptions on the number of contributing 
resonances. Two classes of results were found (BnGa2011-01 and BnGa2011-02). 
The solution BnGa2011-01 gave a helicity amplitude 
$A_{1/2}=0.028\pm0.005\,$GeV$^{-1/2}$ and a branching ratio $\BR(N\eta)=0.16\pm 0.05$, solution BnGa2011-02 
a helicity amplitude $A_{1/2}=0.038\pm0.005\,$GeV$^{-1/2}$ and a branching ratio $\BR(N\eta)=0.21\pm 0.02$. 

With the new data, solutions with two $J^P=5/2^+$ resonance poles only became 
significantly worse and fits with this hypothesis were no longer included in the calculation
of averages 
This leads to an increase of the helicity amplitude 
and the $N\eta$ branching ratio to $A_{1/2}=0.036\pm0.005\,$GeV$^{-1/2}$ and $\BR(N\eta)=0.22\pm 0.04$, respectively if only the solutions with 3 poles in 
BnGa2011-01 and BnGa2011-02~\cite{Anisovich:2011fc} are considered. 
Based on the fits to the 
new data, an additional increase of $A_{1/2}\sqrt{\BR(N\eta)}$ by about 9\% is 
observed while the values for $A_{1/2}$ and $\BR(N\eta)$ optimize at $A_{1/2}=0.032\pm0.006\,$GeV$^{-1/2}$ 
and $\BR(N\eta)=0.33\pm 0.04$. 

The change of the result on the $N\eta$ branching ratio of $N(1650)1/2^-$ is 
hence due to the facts that one set of solutions is discredited by new data, 
that $A_{1/2}\sqrt{\BR(N\eta)}$ is slightly increased and that the fit optimizes at 
a slightly lower helicity coupling.

In 2012, Shklyar {\it et al.} \cite{Shklyar:2012js} published 
the results of a coupled-channel analysis of a large number of
reactions (of course not yet including the new observables becoming available only recently). The authors gave a branching ratio of $0.01\pm0.02$, which we read as $<0.03$ and
used a helicity coupling of $0.063\pm0.007$. 
If our value on $A_{1/2}$ is used, the upper limit 
increases to $<0.12$. It is still incompatible with our finding but the discrepancy is 
reduced. Similar arguments hold true for the results presented in~\cite{Penner:2002ma,Penner:2002md}. 

The value for $A_{1/2}\sqrt{\BR(N\eta)}$ found by
$\eta$MAID2017~\cite{Kashevarov:2017kqb} and MAID2018~\cite{MAID2018} is
consistent with our findings. 

With our value for  $A_{1/2}$, the $N\eta$ branching ratio above 1 would be determined from 
the results presented by Tryasuchev \cite{A.Tryasuchev:2014jda}. In \cite{A.Tryasuchev:2014jda} a high intensity is
assigned to $N(1720)3/2^+$, and the fit quality is not absolutely convincing. Thus we do
not believe that this result excludes the branching ratio reported in this letter. 

Interesting results can be expected if the data presented here will be included 
in other analyses as planned, e.g., within the J\"uBo coupled-channel 
analysis~\cite{deborah_priv}.

\section{Summary {\color{magenta}}}
Summarizing, we have determined the polari\-zation observables $E, T, H, P,$ and $G$
for the reaction $\gamma p\to \eta p$ from measurements using a polarized beam 
and a polarized target. 
Further, the new measurements on the differential cross section 
$d\sigma/d\varOmega$~\cite{Kashevarov:2017kqb},
and new data on the beam asymmetry $\varSigma$~\cite{Collins:2017sgu,Farah-2019}, 
on the polarization observables $T,F$~\cite{Akondi:2014ttg} from MAMI, and on $E$ from CLAS~\cite{Senderovich:2015lek} 
were added to the data base~\cite{Anisovich:2011fc}. The new data provide significant new constraints on the 
$\eta$-photoproduction amplitude.
Branching ratios for $N^*\to N\eta$ decays were determined.
The large $N(1650)1/2^-\to N\eta$ branching ratio found is surprising, given the
previously large difference in the $N\eta$ branching ratios of the
$N(1535)1/2^-$\,/ $N(1650)1/2^-$ nucleon-resonance pair, which was extensively discussed in
literature (see~\cite{Krusche:2003ik} for a summary and the examples in the introduction).
In the standard quark model, this has been taken as evidence for a large mixing
of SU(6)$\times$O(3) states (see Review on Quark Models in~\cite{PDG2019}).
Together with the inversion of the relative sign of the electromagnetic couplings 
of the $N(1535)1/2^-$ and the $N(1650)1/2^-$ state for photoproduction off 
the proton and the neutron~\cite{Anisovich:2015tla}, the
interpretation of these states within the quark model will have to be revised.

We thank the technical staff of ELSA and the par\-ti\-ci\-pating
institutions for their invaluable contributions to the success of
the experiment. We acknowledge support from the \textit{Deutsche
Forschungsgemeinschaft} (SFB/TR16 and SFB/TR110), 
the \textit{U.S. Department of energy, Office of Science, Office of Nuclear Physics under Awards No. DE-FG02-92ER40735}, the \textit{Russian Science foundation}, and the  \textit{Schweizerischer Nationalfonds (200020-156983, 132799, 121781, 117601)}.


\begin{thebibliography}{00}
\bibitem{Isgur:1977ef}
  N.~Isgur and G.~Karl,
  Phys.\ Lett.\ B {\bf 72}, 109 (1977).

\bibitem{Capstick:1986bm}
  S.~Capstick and N.~Isgur,
  Phys.\ Rev.\ D {\bf 34}, 2809 (1986).

\bibitem{Glozman:1995fu}
  L.~Y.~Glozman, D.~O.~Riska,
  Phys.\ Rept.\  {\bf 268}, 263 (1996).

\bibitem{Loring:2001kx}
U.~L\"oring {\it et al.}, 
  Eur.\ Phys.\ J.\ A {\bf 10}, 395 (2001).

\bibitem{Edwards:2011jj}
  R.~G.~Edwards {\it et al.}, 
  Phys.\ Rev.\ D {\bf 84}, 074508 (2011).

\bibitem{Nacher:1999vg}
  J.~C.~Nacher {\it et al.}, 
  Nucl.\ Phys.\ A {\bf 678}, 187 (2000).

\bibitem{Kolomeitsev:2003kt}
  E.~E.~Kolomeitsev {\it et al.}, 
  Phys.\ Lett.\ B {\bf 585}, 243 (2004).

\bibitem{Bruns:2010sv} 
  P.~C.~Bruns, M.~Mai and U.~G.~Meissner,
  Phys.\ Lett.\ B {\bf 697}, 254 (2011)

\bibitem{Mai:2012wy}
  M.~Mai, P.~C.~Bruns and U.-G.~Meissner,
  Phys.\ Rev.\ D {\bf 86}, 094033 (2012).

\bibitem{PDG2010}
K.~Nakamura  {\it et al.} (PDG), JPG {\bf 37}, 075021 (2010).

\bibitem{Glozman:1995tb}
  L.Y.~Glozman, D.O.~Riska,
  Phys.\,Lett.\,B\,{\bf 366},\,305\,(1996).

\bibitem{Zou:2007mk}
  B.~S.~Zou,
  Eur.\ Phys.\ J.\ A {\bf 35}, 325 (2008).

\bibitem{Kaiser:1995cy}
  N.~Kaiser {\it et al.}, 
  Phys.\ Lett.\ B {\bf 362}, 23 (1995).

\bibitem{Vrana:1999nt} 
  T.~P.~Vrana, S.~A.~Dytman, and T.~S.~H.~Lee,
  Phys.\ Rept.\  {\bf 328}, 181 (2000).

\bibitem{Penner:2002ma} 
  G.~Penner and U.~Mosel,
  Phys.\ Rev.\ C {\bf 66}, 055211 (2002).

\bibitem{Penner:2002md} 
  G.~Penner and U.~Mosel,
  Phys.\ Rev.\ C {\bf 66}, 055212 (2002).

\bibitem{Shklyar:2012js} V.~Shklyar H.~Lenske, and U.~Mosel,
  Phys.\ Rev.\ C {\bf 87}, 015201 (2013).

\bibitem{Anisovich:2011fc}
  A.~V.~Anisovich  {\it et al.}, Eur.\ Phys.\ J.\ A\ {\bf 48}, 15 (2012).

\bibitem{Thoma:2007bm} 
  U.~Thoma {\it et al.},
  Phys.\ Lett.\ B {\bf 659}, 87 (2008).

\bibitem{Kashevarov:2017kqb} 
  V.~L.~Kashevarov {\it et al.} [A2 Collaboration],
  Phys.\ Rev.\ Lett.\  {\bf 118}, no. 21, 212001 (2017).

\bibitem{Akondi:2014ttg}
  C.~S.~Akondi {\it et al.},
  Phys.\ Rev.\ Lett.\  {\bf 113}, 102001 (2014).

\bibitem{Senderovich:2015lek}
  I.~Senderovich {\it et al.}, 
  Phys.\ Lett.\ B {\bf 775}, 64 (2016).

\bibitem{Crede:2003ax}
  V.~Crede {\it et al.},
  Phys.\ Rev.\ Lett.\  {\bf 94} 012004 (2005),
  Phys.\ Rev.\ C {\bf 80}, 055202 (2009).

\bibitem{Bartalini:2007fg} 
  O.~Bartalini {\it et al.} [GRAAL Collaboration],
  Eur.\ Phys.\ J.\ A {\bf 33}, 169 (2007)

\bibitem{Williams:2009yj} 
  M.~Williams {\it et al.} [CLAS Collaboration],
  Phys.\ Rev.\ C {\bf 80}, 045213 (2009)

\bibitem{MAID2018}
  L.~Tiator {\it et al.},
  Eur.\ Phys.\ J.\ A\ {\bf 54}, no. 54, 210 (2018).

\bibitem{Shrestha:2012ep} 
  M.~Shrestha and D.~M.~Manley,
  Phys.\ Rev.\ C {\bf 86}, 055203 (2012).

\bibitem{A.Tryasuchev:2014jda}
  V.~A. Tryasuchev,
  Eur.\ Phys.\ J.\ A {\bf 50}, 120 (2014).

\bibitem{PDG2014}
K.~A.~Olive {\it et al.},
Chin.\ Phys.\ C {\bf 38}, 090001 (2014).

\bibitem{PDG2017}
C. Patrignani et al. (Particle Data Group), 
Chin. Phys. C, 40, 100001 (2016) and 2017 update.

 \bibitem{PDG2019}
M. Tanabashi et al. (Particle Data Group), 
Phys. Rev. D 98, 030001 (2018) and 2019 update. 

\bibitem{deborah} 
  D.~R\"onchen {\it et al.}, 
  Eur.\ Phys.\ J.\ A  {\bf 51} 70 (2015).

\bibitem{Gottschall:2013uha}
  M.~Gottschall {\it et al.},
  Phys.\ Rev.\ Lett.\  {\bf 112}, 012003 (2014).

\bibitem{Gottschall:2019pwo} 
  M.~Gottschall {\it et al.} [CBELSA/TAPS Collaboration],
  arXiv:1904.12560 [nucl-ex].

\bibitem{Hartmann:2014mya}
  J.~Hartmann {\it et al.},
  Phys.\ Rev.\ Lett.\  {\bf 113}, 062001 (2014).

\bibitem{Hartmann:2015kpa} 
  J.~Hartmann {\it et al.} [CBELSA/TAPS Collaboration],
  Phys.\ Lett.\ B {\bf 748}, 212 (2015)

\bibitem{Thiel:2012yj}
  A.~Thiel {\it et al.}, Phys.\ Rev.\ Lett.\ {\bf 109}, 102001 (2012).

\bibitem{Thiel:2016chx}
  A.~Thiel {\it et al.}, arXiv:1604.02922 [nucl-ex] (2016).

\bibitem{Hillert:2006yb}
  W.~Hillert, Eur.\ Phys.\ J.\ A\ {\bf 28S1}, 139 (2006).

\bibitem{Dutz:2004eb}
  H.~Dutz, Nucl.\ Instrum.\ Meth.\ A {\bf 526}, 117 (2004).

\bibitem{Suft:2005cq}
  G.~Suft {\it et al.}, Nucl.\ Instrum.\ Meth.\ A\ {\bf 538}, 416 (2005).

\bibitem{CB} E. Aker {\it et al.}, Nucl.\ Instrum.\ Meth.\ A\ {\bf 321}, 69 (1992).

\bibitem{TAPS} R. Novotny, IEEE Trans.\ Nucl.\ Sci.\ {\bf 38}, 379 (1991).

\bibitem{tobias-jan_in_prep}
T.~Seifen, J.~Hartmann {\it et al.}, in preparation.

\bibitem{olsen_maximon} H.~Olsen and L.~C.~Maximon, Phys.\ Rev.\ {\bf 114}, 887 (1959).

\bibitem{Elsner:2008sn}
  D.~Elsner {\it et al.},
  Eur.\ Phys.\ J.\  A {\bf 39}, 373 (2009).

\bibitem{anb}
  F.A.~Natter {\it et al.}, 
  Nucl.\ Instrum.\ Meth.\ B\ {\bf 221}, 465 (2004).

\bibitem{reicherz_nmr} G.~Reicherz {\it et al.}, Nucl.\ Instrum.\ Meth.\ A\ {\bf 356}, 74 (1995).

\bibitem{Annand:2016ppc} 
  J.~R.~M.~Annand {\it et al.} [A2 and MAMI Collaborations],
  Phys.\ Rev.\ C {\bf 93}, no. 5, 055209 (2016).
  doi:10.1103/PhysRevC.93.055209

\bibitem{Drechsel:2007if}
  D.~Drechsel {\it et al.}, 
  Eur.\ Phys.\ J.\ A\ {\bf 34}, 69 (2007).



\bibitem{SAID-GE09}
SAID: http://gwdac.phys.gwu.edu/
%


\bibitem{Collins:2017sgu} 
  P.~Collins {\it et al.},
  Phys.\ Lett.\ B {\bf 771}, 213 (2017)


\bibitem{Anisovich:2004zz} 
  A.~Anisovich {\it et al.}, 
  Eur.\ Phys.\ J.\ A {\bf 24}, 111 (2005).

\bibitem{Anisovich:2006bc} 
  A.~V.~Anisovich and A.~V.~Sarantsev,
  Eur.\ Phys.\ J.\ A {\bf 30}, 427 (2006). 

\bibitem{Anisovich:2007zz} 
  A.~V.~Anisovich {\it et al.}, 
  Eur.\ Phys.\ J.\ A {\bf 34}, 129 (2007).  
\bibitem{Richards:1970cy} 
  W.~B.~Richards {\it et al.},
  Phys.\ Rev.\ D {\bf 1}, 10 (1970).
\bibitem{Brown:1979ii} 
  R.~M.~Brown {\it et al.},
  Nucl.\ Phys.\ B {\bf 153}, 89 (1979).
\bibitem{Prakhov:2005qb} 
  S.~Prakhov {\it et al.},
  Phys.\ Rev.\ C {\bf 72}, 015203 (2005).

\bibitem{Sokhoyan:2015fra}
  V.~Sokhoyan {\it et al.},
 Eur.\ Phys.\ J.\ A {\bf 51}, no. 8, 95 (2015),
  [Eur.\ Phys.\ J.\ A {\bf 51}, no. 12, 187 (2015)].

\bibitem{Gutz:2014wit}
  E.~Gutz {\it et al.} 
  Eur.\ Phys.\ J.\ A {\bf 50} 74 (2014).


\bibitem{Farah-2019}
F.~Afzal {\it et al.} [CBELSA/TAPS Collaboration], 
``Precise beam asymmetry $\Sigma$ data for $\gamma p\to p\eta$ in the 
$p\eta^\prime$-threshold region'', in preparation

\bibitem{Anisovich:2017pox} 
  A.~V.~Anisovich {\it et al.},
  Phys.\ Lett.\ B {\bf 772}, 247 (2017).

\bibitem{pwa-webseite}
https://pwa.hiskp.uni-bonn.de/

%
\bibitem{Krusche:2003ik}
  B.~Krusche {\it et al.}, 
  Prog.\ Part.\ Nucl.\ Phys.\  {\bf 51}, 399 (2003).


\bibitem{Anisovich:2015tla}
  A.~V.~Anisovich {\it et al.}, 
 Eur.\ Phys.\ J.\ A {\bf 51}, no. 6, 72 (2015).

\bibitem{Ronchen:2014cna} 
  D.~R\"onchen {\it et al.},
  Eur.\ Phys.\ J.\ A {\bf 50}, no. 6, 101 (2014)
  Erratum: [Eur.\ Phys.\ J.\ A {\bf 51}, no. 5, 63 (2015)]

\bibitem{deborah_priv}
D.~R\"onchen, private communication 

\end{thebibliography}
\end{document}